\documentclass[12pt]{amsart}
   \usepackage{oldlfont}
   \newtheorem{lemma}{Lemma}[section]
   \newtheorem{theorem}[lemma]{Theorem}

   \newtheorem{definition}[lemma]{Definition}
   \newcommand{\eps}{\varepsilon}
   \newcommand{\Wsob}{\smash{{\stackrel{\circ}{W}}}_2^1(D)}

   \newcommand{\PX}{{\Bbb{P}}}

\renewcommand{\a}{\alpha}

\newcommand{\D}{\Delta}
\newcommand{\p}{\partial}
\renewcommand{\phi}{\varphi}

\newcommand{\R}{{\mathbb R}}

\renewcommand{\k}{\kappa}

\title[Stochastic Thermohaline Circulation]
{Dynamics of the Thermohaline Circulation Under Uncertainty}

\thanks{This work was partly supported by the NSF Grant DMS-0209326,
the National Natural Science Foundation of China
 (No.10171044),  the Natural Science Foundation of Jiangsu Province
  (No.BK2001024), and the Foundation for University Key Teachers of  the
  Ministry of Education of China. }

\author{Wei Wang }
\address[W. Wang ]
{Department of Mathematics\\
 Nanjing University \\
 Nanjing 210008,  China }
\email[W. Wang]{wangshengtaoo@sina.com  }

\author{Jianhua  Sun }
\address[J. Sun ]
{Department of Mathematics\\
 Nanjing University \\
 Nanjing 210008,  China }
\email[J. Sun]{jhsun@nju.edu.cn }

\author{Jinqiao Duan}
\address[J. Duan]
{Department of Applied Mathematics\\
 Illinois Institute of Technology\\
   Chicago, IL 60616, USA }
\email[J.~Duan]{duan@iit.edu}


\begin{document}

\begin{abstract}
The  ocean  thermohaline circulation   under uncertainty is
investigated  by  a random dynamical systems approach. It is shown
that the asymptotic dynamics of  the thermohaline circulation is
described by a
random attractor and by a system with  finite degrees of freedom.
\end{abstract}

\maketitle

\section{Introduction}

The ocean thermohaline circulation(THC) involves water masses
sinking at high latitudes and upwelling at lower latitudes. The
process is maintained by water density contrasts in the ocean,
which themselves are created by atmospheric forcing, namely, heat
and freshwater exchange via evaporation and precipitation at the
air-sea interface.  Thus the ocean THC is driven by fluxes of heat
and freshwater through the air-sea interface.
During the THC, water masses carry heat or cold around the globe.
Thus, it is believed that the global ocean THC plays an important
role in the climate \cite{Siedler}.

The formulation and analysis of mathematical models is
central to the progress of better understanding of the THC dynamics
and its impact on climate change.
Apart from a detailed modeling of the climate system using coupled
general circulation models, sometimes   simplified climate models
could give insight into the general characteristics of the climate
system. The most simplified climate models can be described in terms
of stochastic differential equations \cite{Hass, Cessi0, Lohm}. These
stochastic climate models can be viewed as comprehensive paradigms or
metaphors for particular features of the climate system.

We consider a two-dimensional thermohaline ocean circulation model
in the latitude-depth (meridional) plane, in terms of
the stochastic Navier-Stokes fluid equations (vorticity form) and
the transport equations for heat and salinity, together with air-sea flux or
 Neumann boundary conditions. The noise in the Navier-Stokes   equations
is due to various fluctuations such as random wind stress forcing.

 We intend to
investigate the characteristics of the THC's dynamics when some random effect
is taken into account. Our approach here is a random dynamical systems approach \cite{Arn98}.

This paper is organized as follows. In the next section
we present the THC model, and discuss the well-posedness of this
model in Section 3. Section 4 is devoted to the investigation of the dynamical
behavior of this model: random attractor and finite dimensionality.


\section{A Model for the Thermohaline Circulation }

We consider the ocean  thermohaline circulation in a bounded domain, i.e.,
 a square
$$
D=\{(y,z):\; -l \leq y \leq l,\; 0 \leq z \leq d\},
$$
on the meridional, latitude-depth $(y,z)$-plane, as used by various
authors \cite{Quon, Quon2,  Thual, Cessi,DijkBook}. It is composed
of the Boussinesq version of the Navier-Stokes
equations for oceanic fluid velocity
$(v(y,z,t), w(y,z,t))$
and transport equations for the oceanic salinity $S(y, z, t)$
and the oceanic temperature $T(y, z, t)$ in dimensional form:
\begin{eqnarray*}
v_t+vv_y+wv_z & = & -p_y +\nu \D  v  +\mbox{noise}\; ,\\
w_t+vw_y+ww_z & = & -p_z - g(\a_S S-\a_T T)+\nu \D  w
 +\mbox{noise}\; ,\\
v_y+w_z  & = & 0,\\
T_t+vT_y+wT_z  & = & \k_T  \D  T,\\
S_t+vS_y+wS_z   & = & \k_S  \D  S,
\end{eqnarray*}
where $\a_S$ and $\a_T$ are the coefficients of volume expansion for
salt and heat, respectively; $g$ is the gravitational acceleration;
$\nu$ is the viscosity; and $\k_S$ and $\k_T$ are salt and heat
diffusivities, respectively. The density  is $\rho=\rho_0
(1+\a_S S-\a_T T)$ with $\rho_0$ the mean sea water density.
The noise in the Navier-Stokes   equations
is due to various fluctuations such as random wind stress
forcing \cite{PeiOor92, Washington, FM}.
Presumably,  the noise also affects the transport of heat and salinity
to some extent, but we will ignore this effect.

As discussed in \cite{Thual, Quon}, this may be regarded as a zonally
averaged model of the world ocean. The effect of the rotation can be
parameterized in the magnitude of the viscosity and diffusivity terms.
Introducing the stream function $\psi(y,z,t)$ for the velocity field,
$$v=-\psi_z, \;\; w=\psi_y,$$
we can rewrite the above model in the vorticity form with only three
unknowns $\psi, T, S$:
\begin{eqnarray}
\D\psi_t + J(\psi, \D \psi) &  = & g(\a_T T_y-\a_S S_y)
 + \nu \D^2 \psi  +\dot{\mathcal W}_1\; ,
\label{eqn1}\\
T_t+J(\psi,T) &  = & \k_T \D T,
\label{eqn2}\\
S_t+ J(\psi, S) &  = &\k_S  \D  S,
\label{eqn3}
\end{eqnarray}
where $\mathcal{W}$$_1(y,z,t)$ is a Wiener process defined on a underlying probability
space $(\Omega, {\cal F}, \PX)$ to be specified below.
The noise is described by the generalized time-derivative of the Wiener
 process. The fluctuating noise in the oceanic fluid equation is usually of
a shorter time scale than the response time scale of the large
scale oceanic THC. We thus assume the noise is white in time
(uncorrelated in time)  but
it is allowed to be colored in space, i.e., it may be correlated
in space variables.
Note that the pressure field $p$ is eliminated in the vorticity form of the
Navier-Stokes equations.

Boundary conditions for the oceanic fluid are no normal flow and free-slip
 (no-stress) on the whole boundary \cite{Ped96}:
\begin{eqnarray}
\psi= \D \psi =0 \;\; \mbox{on $\p D$}.
\end{eqnarray}
The boundary conditions  for  temperature and salinity are Neumann type.
 On the air-sea interface $z=d$, the
heat/temperature flux  and  freshawater/salinity flux are prescribed as
\begin{equation}
T_z(y,d) = \lambda (\theta(y)-T), \;\;\;  S_z(y, d)=F(y).
\label{randombc}
\end{equation}
Here $\theta(y)$ is the prescribed (known) atmosphere surface
temperature and $F(y)$ is the mean freshwater flux (known).
Moreover, $\lambda = \frac{B_T}{\rho_0 C_p \k_T}$, with $B_T$
being the surface exchange coefficient of heat and $C_p$ the
heat capacity.

Zero flux boundary conditions are taken for $T$ and $S$ on fluid
bottom $(z=0)$ and on fluid side $(y=\pm l)$:
\begin{eqnarray}
T_z(y,0)=0, \; S_z(y,0)=0,\; \\
T_y(\pm l, z)=0, \; S_y(\pm l, z)=0.
\end{eqnarray}

The THC model above involves stochastic and deterministic partial differential equations (PDEs) and Neumann boundary conditions.

\section{Well-Posedness}

In this section we will show that (1)-(3) defines a well-posed
model. First we introduce some function spaces from the theory
of partial differential equations.

Let $W_2^1(D)$ be the Sobolev space of function on $D$ with the
first generalized derivative in $L_2(D)$, the function space of
square integrable functions on $D$ with norm and inner product
$$||u||_{L_2}=(\int_D|u(x)|^2dD)^{\frac{1}{2}},\;\;\;
(u,v)_{L_2}=\int_Du(x)v(x)dD, \;\;\; u,v\in L_2(D)$$
The space $W_2^1(D)$ is equipped with the norm
$$||u||_{W_2^1}=||u||_{L_2}+||\partial_y
u||_{L_2}+||\partial_z u||_{L_2}$$
Let the   $\Wsob$ be the space of functions vanishing
on the boundary $\partial D$ of $D$.
The norm of this space is defined as
\begin{equation}\label{}
\|u\|_{\Wsob}=\|\partial_y u\|_{L_2}+\|\partial_z u\|_{L_2}.
\end{equation}

Similarly, we can define function spaces on the interval $(0,d)$
denoted by $L_2(0,d)$ and $W_2^1(0,d)$.

Another Sobolev space is     $\dot{W}^1_2(D)$ which is a subspace
of $W_2^1(D)$ consisting of functions $u$ with zero mean:
$\int_DudD=0$. A norm equivalent to the $W_2^1$-norm on
$\dot{W}^1_2(D)$ is given by the right hand side of (8). For
functions in $L_2(D)$ having the same property, we write as
$\dot{L}_2(D)$.

There exists a {\em continuous trace operator}:
\begin{equation*}
\gamma_{\partial D}: W_2^1(D)\to H^\frac{1}{2}(\partial D).
\end{equation*}
Here $H^\frac{1}{2}(\partial D)$ is a boundary space, see Adams \cite{Ada75}
Chapter 7 or below. Similarly, we can introduce trace operators that
map onto a part of the boundary of $\partial D$ for instance for the subset
$\{(y,z)\in \bar D|z=d\}$ of $\bar D$. For this mapping we will write
\begin{equation}\label{trace1}
\gamma_{z=d}: W_2^1(D)\to H^\frac{1}{2}(0,d).
\end{equation}
The adjoint operator
\[\gamma_{z=d}^\ast:(H^\frac{1}{2}(0,d))^\prime\to ({W}_2^1(D))^\prime\]
is also continuous, where $^\prime$ denotes the dual space for a given
Banach space.

If set vorticity $ q=\D \psi$ we can homogenize boundary conditions
 to obtain:
\begin{eqnarray}
q_t + J(\psi, q) &  = & g(\a_T T_y-\a_S S_y)
   + \nu \D q  +\dot{\mathcal{W}}_1\; ,
\label{new1}\\
T_t+J(\psi,T) &  = & \k_T \D T
  + \gamma_{z=d}^\ast (\lambda (\theta(y)-\gamma_{z=d}T) ),
\label{new2}\\
S_t+ J(\psi, S) &  = &\k_S  \D  S
  + \gamma_{z=d}^\ast  F(y) .
\label{new3}
\end{eqnarray}

New  homogeneous boundary conditions are:
\begin{eqnarray*}
\psi=0,  \;\;\;  q =0 \;\; \mbox{on $\p D$}, \\
T_z(y,d)  = 0, \;\;\;   S_z(y, d)=0, \\
T_z(y,0)=0,     \;\;\;    S_z(y,0)=0,\; \\
T_y(\pm l, z)=0, \;\;\; S_y(\pm l, z)=0.
\end{eqnarray*}\\


For convenience, we introduce the vector notation for unknown
geophysical quantities
\begin{equation}\label{}
u=(q,T,S).
\end{equation}

Now we can define the linear differential operator from (10)-(12)

\begin{equation*}
{A}u =\left(
\begin{array}{l}
-{\nu} \Delta q\\
-\kappa_T \Delta T\\
-\kappa_S \Delta S
\end{array}
\right).
\end{equation*}

We assume that $F$ and $\theta\in L_2(0,d)$. Note that
$$\frac{d}{dt}\int_DSdydz=\int^d_0F(y)dy=constant.$$
It is reasonable (see \cite{DijkBook}) to assume that
$$\int_0^dF(y)=0$$
and thus $\int_DSdydz$ is constant in time and we may assume it is
zero:
\begin{equation*}
\int_DSdydz = 0.
\end{equation*}
Thus we have the usual Poincar\'e inequality for $S$. However,
this is not true for $T$. Fortunately we can derive the following
Poincar\'e inequality for $T$
\begin{equation}
\label{} \|T\|^2 \le c(\Omega) (\|\gamma_{z=1}T\|_{L^2}^2 +
\|\nabla T\|^2_{L_2}),
\end{equation}
as in Temam \cite{Tem97} where $c(\Omega)$ is a constant dependent
on $\Omega$.

 Introduce the phase space for our system
 $H=L_2(D)\times L_2(D)\times \dot{L}_2(D)$ with the usual $L_2$ inner
product and $V=\Wsob\times W_2^1(D)\times \dot{W}_2^1(D)$.

It is obvious that the linear operator $A:V\to V^\prime$ is
positive definite. And define the nonlinear operator
$G(u):=G_1(u)+G_2(u)$ where
$$ G_1(u)[y,z]= \left(
\begin{array}{c}
-J(\psi,q)\\
-J(\psi,T)\\
-J(\psi,S)
\end{array}
\right)[y,z]. $$
and
$$G_2(u)[y,z]= \left(
\begin{array}{l}
g(\alpha_T T_y-\alpha_S S_y)\\
\gamma_{z=d}^\ast(\lambda(\theta(y)-\gamma_{z=d}T)) \\
\gamma_{z=d}^\ast F(y)
\end{array}
\right)[y,z]. $$

Then the THC system can be rewritten as a stochastic differential
equation on $V^\prime$:
\begin{equation}
\label{} \frac{du}{dt}+Au=G(u)+\dot{W} \qquad u(0)=u_0\in H,
\end{equation}
where $W=(\mathcal{W}$$_1,0,0)$, $\dot{W}$ is a white noise as the
generalized temporal derivative of a Wiener process $W$ with
continuous trajectories on $\mathbb{R}$ and with values in
$L_2(D)$. It is sufficient for this regularity that the trace of
the covariance is finite with respect to the space $L_2(D)$: ${\rm
tr}_{L_2}Q< \infty$. In particular, we can choose the canonical
probability space $(\Omega,\mathcal{F},\mathbb{P})$ where the set
of elementary events $\Omega$ consists of the paths of $W$ and the
probability measure $\mathbb{P}$ is the Wiener measure with
respect to covariance $Q$.

Through integration by parts and direct estimation or from \cite{ChuDuaSchm01a} we   have the following lemmas.

\begin{lemma}
The operator $G_1:V\to H$ is continuous. In particular, we have
$$ \langle G_1(u),u\rangle=0.$$
\end{lemma}
and

\begin{lemma}
The following estimation  holds
 $$ \|G_2(u)\|_{V^\prime}\le c_1\|u\|_V+c_2.$$
for some  positive constants $c_1,\,c_2$.
\end{lemma}

In the following we need a stationary Ornstein-Uhlenbeck process
solving the linear stochastic equation on $D$
\begin{equation}
\frac{d\eta}{dt}-\nu(k+1)\D \eta = \dot{\mathcal{W}}_1
\end{equation}
with the homogeneous Neumann boundary condition at $\partial D$.
Here $k$ can be seen as a very large controlling parameter.

\begin{lemma}
Suppose that the covariance $Q$ has a finite trace : ${\rm
tr}_{L_2} Q<\infty$. Then (16) has a unique stationary solution
generated by
 $$ (t,\omega)\to \eta(\theta_t\omega).$$
Moreover, $\eta$ is a tempered random variable in $W_2^1(D)$ and
has trajectories in the space $L^2_{loc}(\R;W_2^3(D))$. Then
$Z(\omega)=(\eta(\omega),0,0)$ is a random variable in $V$.
\end{lemma}

For the proof we refer to \cite{DaPrato}.\\

If we set
\begin{equation}
(\tilde{q}, T, S)=v:=u-Z=(q-\eta, T, S),
\end{equation}
then we obtain a random differential equation in $V^\prime$
\begin{equation} \label{rds}
\frac{dv}{dt}+Av=G_1(v)+\tilde{G}_2(v+Z(\theta_t\omega)),\qquad
v(0)=v_0\in H,
\end{equation}
where $\tilde{G}_2=G_2+(J(\eta,\tilde{q})+J(\D^{-1}
\tilde{q},\D\eta)+J(\eta, \D\eta)-\nu k\D\eta,0,0)$.\\

The above equation (\ref{rds}) is a differential equation with
random coefficients  then it can be treated sample-wise for any
sample $\omega$. We are looking for solution $v$ in
\begin{equation*}
 C([0, \tau]; H)\cap L_2(0, \tau; V),
\end{equation*}
for all $\tau>0$. If we can solve this equation then $u:=v+Z$
defines a solution version of (15). For the well posedness of the
problem we now have the following result.

\begin{theorem}
({\bf Well-Posedness}) For any time $\tau>0$, there exists a
unique solution of (\ref{rds}) in $C([0,\tau];H)\cap L_2(0,\tau;V)$. In
particular, the solution mapping
$${{\mathbb{R}}}^+\times \Omega\times H\ni(t,\omega,v_0)\to v(t)\in H
$$
is measurable in its arguments and the solution mapping $H\ni
v_0\to v(t)\in H$ is continuous.
\end{theorem}
\begin{proof}
By the properties of $A$ and $G_1$ (see Lemma 3.1), the random
differential equation (18) is essentially similar to the
$2$-dimensional Navier Stokes equation. Note that $\tilde{G}_2$ is
only an affine mapping. Hence we have existence and uniqueness and
the above regularity assertions.
\end{proof}

Now we can define a random dynamical system since the solution
mapping
$$ {\mathbb{R}}^+\times\Omega\times H\ni (t,\omega,v_0)\to
v(t,\omega,v_0)=:\phi(t,\omega,v_0)\in H. $$
is well defined. First we define a so-called metric dynamical system
$(\Omega,\mathcal{F},\mathbb{P}$,\\
$(\theta_t)_{t\in\R}).$ $ {\theta_t: \Omega \rightarrow \Omega,
t\in \R}$ is a family of measure preserving transformations such that
$(t, \omega)\mapsto \theta_t \omega$ is measurable, $\theta_0=id,
\theta_{t+s}=\theta_{t}\theta_{s},$ for all $s, t \in \R$.
Furthermore, the shift $\theta_t$ is ergodic if we define it as
$$w(\cdot,\theta_t\omega)=w(\cdot+t,\omega)-w(t,\omega)\quad
\text{for }t\in{\mathbb{R}}$$
which is called the {\em Wiener shift}. A random dynamical
system $\phi(t,\omega,u)$ is well defined now since the cocycle
property
\begin{align*}
&\phi(t+\tau,\omega,u)=\phi(t,\theta_\tau\omega,\phi(\tau,\omega,u))
\quad\text{for }t,\,\tau\ge 0\\
&\phi(0,\omega,u)=u
\end{align*}
for any $\omega\in\Omega$ and $u\in H$.  For more detail about
random dynamical systems we refer to \cite{Arn98}. \\

\section{Random dynamics}

In this section, we investigate random dynamics of the THC.
First we are going to show that the random THC model
is dissipative, i.e., it has an random absorbing set in the following
sense:
\begin{definition}
A random set $B=\{B(\omega)\}_{\omega\in \Omega}$ consisting of
closed bounded sets $B(\omega)$ is called absorbing for a random
dynamical system $\phi$ if we have for any random set
$D=\{D(\omega)\}_{\omega\in\Omega},\,D(\omega)\in H$ bounded,
such that $t\to \sup_{y\in D(\theta_t\omega)}\|y\|_H$ has a subexponential
growth for $t\to\pm\infty$
\begin{align}
\begin{split}
&\phi(t,\omega,D(\omega))\subset B(\theta_t\omega) \quad\text{for }t
\ge t_0(D,\omega)\\
&\phi(t,\theta_{-t}\omega,D(\theta_{-t}\omega))\subset
B(\omega)\quad\text{for }t\ge t_0(D,\omega).
\end{split}
\end{align}
$B$ is called forward invariant if
\[
\phi(t,\omega,u_0)\in B(\theta_t\omega)\quad \text{if }  u_0\in
B(\omega)\quad \text{for }t\ge 0.
\]
\end{definition}

 Consider THC system separately. For this we introduce the spaces
\begin{align*}
\tilde H&= L_2(D)\times\dot{L}_2(D),\\
\tilde V&= W_2^1(D)\times \dot{W}_2^1(D).
\end{align*}
We choose a subset of dynamical variables of our system
(10)-(12)
\begin{equation}
\tilde v=(T,S).
\end{equation}
Applying the chain rule to $\|\tilde v\|_H^2$, we obtain by Lemma 3.1
\begin{align}
\begin{split}
\frac{d}{dt} &\|\tilde v\|_{\tilde H}^2+2\kappa_T\|\nabla
 T\|_{L_2}^2 +2\kappa_S\|\nabla S\|_{L_2}^2\\
=&2\lambda(\theta(y),\gamma_{z=d}T)_{L_2}-2\lambda(\gamma_{z=d}T,
\gamma_{z=d}T)_{L_2}+2(F(y),\gamma_{z=d}S)_{L_2}.
\end{split}
\end{align}

Cauchy-Schwarz inequality yields the following estimates
\begin{align*}
2\lambda(\theta(y),\gamma_{z=d}T)_{L_2}
-2\lambda(\gamma_{z=d}T,\gamma_{z=d}T)_{L_2}
\le \frac{\lambda}{a}\|\theta(y)\|^2_{L_2}+(a-2)\lambda
\|\gamma_{z=d}T \|^2_{L_2},
\end{align*}
where $a$ is a positive constant that satisfies
$2>a>2-\frac{2\kappa_T}{\lambda}$. Applying the trace theorem
$\|\gamma_{z=d}S\|^2_{L_2}\le c_3\|\gamma_{z=d}S\|_{H^\frac{1}{2}}
\le c_3\|S\|_{W_2^1}$ and for any $\eps>0$ we can find a $c_4(\eps)$
such that
\begin{align*}
2(F(y),\gamma_{z=d}S)_{L_2} \le
\eps\|S\|^2_{W_2^1}+c_4(\eps)\|F(y)\|^2_{L_2},
\end{align*}
for some $\eps$.

Thus by using the Poincar\'e inequality for $S\in \dot{W}_2^1(D)$,
(14) and choosing $\eps$ small enough, we conclude

\begin{align}
\begin{split}
\frac{d}{dt} \|\tilde v\|_{\tilde H}^2+\alpha(\|\nabla
\tilde v\|_{\tilde H}^2+\|\tilde v\|^2_{\tilde{H}})
\le c_5
\end{split}
\end{align}
where $\alpha$, $c_5$ is determined by $\lambda$, $\kappa_T$,
$\kappa_S$, $\|\theta\|_{L_2}$, $\|F(y)\|_{L_2}$ and $\nabla
\tilde v$ is defined by $(\nabla_{y,z}T,\nabla_{y,z}S)$.\\

By the Gronwall inequality, we finally conclude that
\begin{equation}
\|\tilde v\|_{\tilde H}^2 \le
\|\tilde v(0)\|_{\tilde H}^2e^{-\alpha t}+\frac{c_{5}}{\alpha}.
\end{equation}
Then we can easily see that the ball $B(0,R_1)$, where
$R_1^2=\frac{2c_{5}}{\alpha}$, absorbs
$\tilde{v}$ in the sense of definition 4.1.\\

To prove the dissipativity of the dynamical system $\varphi$ we
have to study $\|\tilde q\|_{L_2}$. From (18), we have
\begin{equation}
\tilde q_t=-J(\psi,q)+g(\alpha_T T_y-\alpha_S S_y)+\nu\D\tilde
q-\nu k\D\eta.
\end{equation}\\
Then
\begin{align}
\begin{split}
 \frac{1}{2}\frac{d}{dt}\|\tilde q\|^2_{L_2}=&(J(\psi,q),\eta)
 -\nu\|\nabla\tilde q\|^2_{L_2}-\nu k(\D\eta,\tilde q)\\&+g(\alpha_T
T_y-\alpha_S S_y,\tilde q).
\end{split}
\end{align}

From the definition of $J(\cdot.\cdot)$ and the Cauchy-Schwarz inequality
we have
\begin{align*}
\begin{split}
 (J(\psi,q),\eta)&=(J(\psi,\eta),q)\le\|\nabla q\|_{L_2}\|q\|_{L_2}\|
 \nabla\eta\|_{L_2}\\
&\le\frac{\nu}{2}\|\nabla
q\|^2_{L_2}+\frac{1}{2\nu}\|\nabla\eta\|^2_{L_2}\|q\|^2_{L_2}\\
&\le\frac{\nu}{2}\|\tilde
q\|^2_{W_2^1}+\frac{\nu}{2}\|\eta\|^2_{W_2^1}+\frac{1}{2\nu}\|\eta\|^2_
{W_2^1}\|\tilde q\|^2_{L_2}+\frac{c_6}{8\nu}\|\eta\|^4_{W_2^1}.
\end{split}
\end{align*}
Here $c_6$ is the constant in the Poincar\'e inequality for
$\eta\in \stackrel{\circ}{W_2^1}(0,d)$. And for any $\eps>0$
we can find $c_7(\eps)>0$ and $c_8(\eps)>0$ such that
\begin{align*}
\begin{split}
g(\alpha_T T_y-\alpha_S S_y,\tilde q)&\le \tilde g\|\nabla \tilde
v\|_{\tilde H}\|\tilde q\|_{L_2}\\&\le \frac{\eps}{4}\|\tilde
q\|^2_{L_2}+c_7(\eps)\tilde g^2\|\nabla v\|^2_{\tilde H},
\end{split}
\end{align*}

\begin{align*}
\begin{split}
-\nu k(\D\eta,\tilde q)&=\nu k(\nabla \eta, \nabla \tilde
q)\\&\le\frac{\eps}{4\lambda_1}\|\tilde q\|^2_{W_2^1}+\lambda_1
c_8(\eps)k^2\nu^2\|\eta\|^2_{W_2^1},
\end{split}
\end{align*}
where $\tilde g$ is determined by $\alpha_T$, $\alpha_S$ and $g$,
$\lambda_1$ is the first eigenvalue of the operator $-\D$ on
$(0,d)$ with Neumann boundary condition.

Collecting all these estimates, we have
\begin{align*}
\begin{split}
\frac{d}{dt}\|\tilde q\|^2_{L_2}&\le
-\gamma(\theta_t\omega)\|\tilde q\|^2_{L_2}+\delta(\eps)\|\nabla
\tilde v\|^2_{\tilde H}+r(\theta_t\omega)\\
&\le-\gamma(\theta_t\omega)\|\tilde q\|^2_{L_2}+r(\theta_t\omega)+
\frac{\delta(\eps)c_5}{\alpha}-\frac{\delta(\eps)}{\alpha}\frac{d}{dt}
\|\tilde v\|^2_{\tilde H},
\end{split}
\end{align*}
where
$\gamma(\omega)=\lambda_1\nu-\eps-\frac{1}{\nu}\|\eta\|^2_{W_2^1}$,
$\delta(\eps)=c_7(\eps)\tilde g^2$ and

\begin{align}
r(\omega)=(2\lambda_1 c_8(\eps)k^2\nu^2+\nu)\|\eta\|^2_{W_2^1}
 +\frac{c_6}{4\nu}\|\eta\|^4_{W_2^1}.
\end{align}

Then  Gronwall inequality yields
\begin{align}
\begin{split}
 \|\tilde q(t,\omega, u_0)\|^2_{L_2}&\le
 \|v_0\|^2_{H}e^{-\int^t_0\gamma(\theta_s\omega)ds}\\&\;\;\;\;+
 \int_0^t(\frac{\delta(\eps)c_5}{\alpha}+r(\theta_s\omega)-
 \frac{\delta(\eps)}{\alpha}\frac{d}{ds} \|\tilde v(s)\|^2_{\tilde H})
 e^{-\int^t_s\gamma(\theta_\tau\omega)d\tau}ds\\&
 \le\|v_0\|^2_{H}e^{-\int^t_0\gamma(\theta_s\omega)ds}\\&\;\;\;\;+
  \int_0^t(\frac{\delta(\eps)c_5}{\alpha}+r(\theta_s\omega))
 e^{-\int^t_s\gamma(\theta_\tau\omega)d\tau}ds\\&\;\;\;\;+
 \frac{\delta(\eps)}{\alpha}\int^t_0\|\tilde v\|^2_{\tilde
 H}\gamma(\theta_s\omega)e^{-\int^t_s\gamma(\theta_\tau\omega)d\tau}ds+
 \frac{\delta(\eps)}{\alpha}\|\tilde v(0)\|^2_{\tilde H}
 e^{-\int^t_0\gamma(\theta_s\omega)ds}.
\end{split}
\end{align}

We will show  that the right hand of (27) is finite as
$t\to\infty$. In fact we have

\begin{lemma}
If the controlling parameter $k$ is large enough that
$$\lambda_1>\frac{tr_{L_2}Q}{(k+1)\nu^3}$$
and $\eps<\frac{\lambda_1\nu}{2}$ then
$$E\gamma(\omega)>0.$$
\end{lemma}
\begin{proof}
It$\hat{o}$ formula applied to $\|\eta\|^2_{L_2}$ yields
\begin{equation*}
\|\eta(\theta_\tau\omega)\|^2_{L_2}+2(k+1)\nu\int^{\tau}_0\|
\eta(\theta_s\omega)\|^2_{W_2^1}ds
=\|\eta(\omega)\|^2_{L_2}+2\int^{\tau}_0(\eta,dw)_{L_2}+\tau
tr_{L_2}Q.
\end{equation*}
Hence we can easily get that
$E\|\eta\|^2_{W_2^1}\le\frac{\lambda_1\nu^2}{2}$. Then
$E\gamma(\omega)>0.$
\end{proof}

Now we can estimate the $\|\tilde{q}\|^2_{L_2}$. First we have
$$\lim_{t\to\infty}(\|v_0\|^2_{H}+\frac{\delta(\eps)}{\alpha}(c_5t+
\|\tilde{v}(0)\|^2_{\tilde{H}}))e^{-\int^t_0\gamma(\theta_s\omega)ds}=0,
\;\;\;P.a.s.$$ And note that $\|\tilde{v}\|^2_{\tilde{H}}$ is
bounded by $\|\tilde v(0)\|_{\tilde H}^2e^{-\alpha
t}+\frac{c_{5}}{\alpha}$ which tends to $\frac{c_5}{\alpha}$
exponentially. We replace $\omega$ by $\theta_{-t}\omega$ to
construct the radius of the absorbing set. Then we have
\begin{align*}
&\lim_{t\to\infty}\int^t_0(r(\theta_{t-s}\omega)+\frac{\delta(\eps)}{\alpha}
\|\tilde v\|^2_{\tilde
H}\gamma(\theta_{t-s}\omega))e^{-\int^t_s\gamma
(\theta_{t-\tau}\omega)d\tau}ds\\
=&\lim_{t\to\infty}\int^0_{-t}(r(\theta_{s}\omega)+\frac{\delta(\eps)}{\alpha}
\|\tilde v\|^2_{\tilde
H}\gamma(\theta_{s}\omega))e^{-\int^t_s\gamma
(\theta_{\tau}\omega)d\tau}ds\\
=:&R^2_2(\omega)<\infty.
\end{align*}

Collecting all the estimates we have
\begin{lemma}
The random set $\{B(\omega)\}_{\omega\in\Omega}$ given by closed
balls $B(0,R(\omega))$ in $H$ with center zero and radius
$R^2(\omega):=R^2_1+R^2_2(\omega)$ is an absorbing and forward
invariant set   for the random dynamical
system   $\phi$ generated by (\ref{rds}).
\end{lemma}

For the application in the following we need the particular
regularity of the absorbing set.   To this end
we introduce the function space
$$ {\mathcal{H}}^s:=\{u\in H:
\|u\|_s^2:=\|A^\frac{s}{2}u\|_H^2<\infty\}$$ where $s\in
{\mathbb{R}}$. The operator $A^s$ is the $s$-th power of the
positive and symmetric operator $A$. Note that these spaces are
embedded into the Sobolev spaces $H^s,\, s>0$. The norm of
these spaces is denoted by $\|\cdot\|_{H^s}$.\\

For our aim we show that $v(1,\omega,D)$ is a bounded set in
${\mathcal{H}}^s$ for some $s>0$. Consider $t\|v(t)\|^2_s$ and by
chain rule we have
 $$\frac{d}{dt}(t\|v(t)\|_s^2)=\|v(t)\|_s^2+t\frac{d}{dt}\|v(t)\|_s^2.$$
The second term in the above formula can be expressed as follows
\begin{align*}
t\frac{d}{dt}(A^\frac{s}{2}v,A^\frac{s}{2}v)_H=&2t(\frac{d}{dt}v,A^sv)_H
=-2t(Au,A^sv)_H+2t(G_1(v),A^sv)_H\\
&+2t(\tilde{G}_2(v+Z(\theta_t\omega)),A^sv)_H.
\end{align*}
Notice that for $\eps>0$ there are constants $c_9-c_{14}$ such that
$$(J(\eta,\tilde{q}), A^s v)_H\le
c_9\|v\|_V \|\eta\|_{1+s}\|v\|_{1+s}\le
c_{10}(\eps)\|\eta\|_{1+s}\|v\|^2_V+\eps\|v\|^2_{1+s},$$
 $$(J(\psi,\D\eta),A^s
v)_H\le c_{11}\|\D\eta\|_1\|\psi\|_{1+s}\|v\|_{1+s}\le
c_{12}(\eps)\|\eta\|_2^2\|v\|^2_{L_{\infty}(0,T;H)}+\eps\|v\|^2_{1+s},$$
$$(J(\eta,\D\eta), A^sv)_H\le
c_{13}\|\D\eta\|_1\|\eta\|_{1+s}\|v\|_{1+s}\le
c_{14}(\eps)\|\eta\|_2^2\|\eta\|^2_{1+s}+\eps\|v\|^2_{1+s},$$
and
$$\int^t_0\|v\|_s^2ds\le c_{15}\int^t_0\|v\|^2_Vds, \;\;\; s\le 1$$
for the embedding constant $c_{15}$ between ${\mathcal{H}}^s $ and
$V$. Then by the similar arguments as in \cite{ChuDuaSchm01a} or \cite{DuanGaoSchm} we have
the following dissipative property of THC.

\begin{lemma}
For the random dynamical system  $\phi$ generated by (\ref{rds}),
there exists a
compact random  set $B=\{B(\omega)\}_{\omega\in \Omega}$ which
satisfies Definition 4.1.
\end{lemma}

We define
\begin{equation}
B(\omega)=\overline{\phi(1,\theta_{-1}\omega,B(0,R(\theta_{-1}\omega)))}
\subset {\mathcal{H}}^s,\quad 0<s<\frac14.
\end{equation}
In particular, ${\mathcal{H}}^s$ is compactly embedded in $H$. \\

Now we show that the dynamics of THC is described by a random
attractor. First we recall the following basic concept; See
\cite{CrauFlan}.
\begin{definition}
Let $\phi$ be a random  dynamical system. A random set
$A=\{A(\omega)\}_{\omega\in \Omega}$ consisting of compact
nonempty sets $A(\omega)$ is called random global attractor if for
any random bounded set $D$ we have for the limit in probability
$${(\mathbb{P})} \lim_{t\to \infty} {\rm dist}_H(\phi(t,\omega,
D(\omega)),A(\theta_t\omega)) = 0$$
and $$\phi(t,\omega,A(\omega))=A(\theta_t\omega)$$
for any $t\ge 0$ and $\omega\in\Omega$.
\end{definition}

The following theorem \cite{CrauFlan} gives a condition of the existence of random attractor.
\begin{theorem}
Let $\phi$ be a random dynamical system on the state space $H$
which is a separable Banach space such that $x\to\phi(t,\omega,x)$
is continuous. Suppose that $B$ is a set ensuring the
dissipativity given in Definition (4.1). In addition, $B$ has a
subexponential growth (see Definition (4.1)) and is regular
(compact). Then the dynamical system $\phi$ has a random
attractor.
\end{theorem}

Then from the above analysis we have the following result to the random
system $\phi$ generated by  (\ref{rds}), and, via the transformation (17), thus to the original
stochastic  THC model.

\begin{theorem} \label{attractor}
({\bf Random Attractor})
 The THC model (1)-(3) has a random attractor
in the phase space $H=L_2(D)\times L_2(D)\times \dot{L}_2(D)$.
\end{theorem}

It is well known that, the global attractor for certain
deterministic infinite-dimensional systems has finite dimension  \cite{Tem97}. This result leads to the fact that
asymptotic behavior of these systems can be described using
finite-dimensional systems. And the similar theory has been
developed for random dynamical systems; see, for instance,
\cite{CraFla99, Deb98, Schm97c}.

However, for   the THC system we will apply another
approach to prove the fact that the random attractor of (1)-(3) has
only finitely many degrees of freedom. Namely we will use the
concept of {\em determining functionals  in probability} as in Chueshov et al. \cite{ChuDuaSchm01a}.
This concept was introduced by Foias
and Prodi \cite{FoiPro67} for deterministic systems.

\begin{definition}
We call a set ${\mathcal{L}}=\{l_j,\;j=1,\cdots, N\}$ of linear
continuous and linearly independent functionals on a space $X$
continuously embedded in $H$ (for instance $X={\mathcal{H}}^s$
or $V$) asymptotically determining in probability if
$$({\mathbb{P}})\lim_{t\to\infty}\int_t^{t+1}
\max_j|l_j(\phi(\tau,\omega,v_1)-\phi(\tau,\omega,v_2))|^2d\tau=0 $$
for two initial conditions $v_1,\,v_2\in H$ implies
$$({\mathbb{P}})\lim_{t\to\infty}\|\phi(t,\omega,v_1)
-\phi(t,\omega,v_2)\|_H=0.$$
\end{definition}

We introduce a constant $\varepsilon_{\mathcal{L}}$ to describe
the qualitative difference of the space $H$ and $X$ for some set of
functionals
\begin{equation}
\|u\|_H\le
C_{{\mathcal{L}}}\max_{l_i\in{\mathcal{L}}}|l_i(u)|
+\eps_{{\mathcal{L}}}\|u\|_X,\quad C_{\mathcal{L}}>0.
\end{equation}
We cite the following theorem from \cite{ChuDuaSchm01a}.

\begin{theorem}
Let ${\mathcal L}=\{ l_j : j= 1,...,N\}$ be a set of linear
continuous and linearly independent functionals on $X$. We assume
that we have an absorbing and forward invariant set $B$ in $X$
such that for $\sup_{v\in B(\omega)}\|v\|_{X}^2$ the expression
$t\to\sup_{v\in B(\theta_t\omega)}\|v\|_{X}^2$ is locally
integrable and subexponentially growing. Suppose there exist a
constant $c_{16}>0$ and a measurable function $l\ge0$ such that
for $v_1,\,v_2\in V$ we have for
$\tilde{G}(\omega,v)=G_1(v)+\tilde{G}_2(v+Z(\omega))$
\begin{align*}
\begin{split}
\langle - A(v_1-v_2)+\tilde{G}(\omega,v_1)-\tilde{G}(\omega,v_2),
v_1-v_2\rangle\\
\le -c_{16}\|v_1-v_2\|_V^2+ l(v_1,v_2,\omega)\|v_1-v_2\|_H^2.
\end{split}
\end{align*}
Assume that
\begin{equation*}
\frac{1}{m}{\mathbb{E}}\left\{\sup_{v_1,v_2\in B(\omega)}\int_0^m
l(\phi(t,\omega,v_1),\phi(t,\omega,v_2),\theta_t\omega)dt\right\}<
c_{16}\eps_{{\mathcal L}}^{-2}
\end{equation*}
for some $m>0$. Then ${\mathcal L}$ is a set of asymptotically
determining functionals in probability for random dynamical system
$\phi$.
\end{theorem}

\bigskip
Now we go back to the  random THC system.
Note that for $\eps>0$ there are constants
$c_{17}(\eps), \cdots, c_{22}(\eps)$ such that
\begin{align*}
\begin{split}
&|\langle J(\tilde{\psi_1},\tilde{q}_1)-J(\tilde{\psi}_2,\tilde{q_2}),
\tilde{q}_1-\tilde{q}_2\rangle|
=|\langle J(\tilde{\psi}_2-\tilde{\psi}_1,\tilde{q}_1), \tilde{q}_2
-\tilde{q}_1\rangle|\\
\le& \frac{\eps}{4}\|\tilde{q}_1-\tilde{q}_2\|_{V}^2+c_{17}(\eps)\|
\tilde{q}_1\|^2_{\Wsob}\|\tilde{q}_1-\tilde{q}_2\|^2_{L_2}.
\end{split}
\end{align*}
Similarly
\begin{align*}
|\langle J(\tilde{\psi_1},T_1)-J(\tilde{\psi}_2,T_2),
T_1-T_2\rangle|
\le\frac{\eps}{4}\|v_1-v_2\|_{V}^2+c_{18}(\eps)\|T_1\|^2_{W_2^1(D)}
\|T_1-T_2\|^2_{L_2},
\end{align*}
\begin{align*}
|\langle J(\tilde{\psi_1},S_1)-J(\tilde{\psi}_2,S_2)S_1-S_2\rangle|
\le\frac{\eps}{4}\|v_1-v_2\|_{V}^2+c_{19}(\eps)\|S_1\|^2_{\dot
W_2^1(D)}\|S_1-S_2\|^2_{L_2},
\end{align*}
\begin{align*}
|\langle \alpha_T T_y-\alpha_S S_y, \tilde{q}_1-\tilde{q}_2\rangle|
\le\frac{\eps}{4}\|v_1-v_2\|_{V}^2+c_{20}(\eps)\|\tilde{q}_1
-\tilde{q}_2\|^2_{L_2},
\end{align*}
and
\begin{align*}
|\langle J(\tilde{\psi_1},\D\eta)-J(\tilde{\psi}_2,\D\eta),
\tilde{q}_1-\tilde{q}_2\rangle| \le
c_{21}\|v_1-v_2\|_{V}^2\|\eta\|^2_{W_2^1}.
\end{align*}
Then we have
\begin{equation*}
\begin{split}
&\langle-A(v_1-v_2)+\tilde{G}(\omega,v_1)-\tilde{G}(\omega,v_2),
v_1-v_2\rangle\\
\le&-c_{22}\|v_1-v_2\|^2_{V}+l(v_1,v_2,\omega)\|v_1-v_2\|^2_{H}.
\end{split}
\end{equation*}
Here we can take $k$ is large enough and $\eps$ is small enough
such that $Ec_{22}=1-\eps-c_{21}E\|\eta\|^2_{W_2^1}>0.$ And
$$l(v_1,v_2,\omega)=c_{17}(\eps)\|\tilde{q}_1\|^2_{\Wsob}+c_{18}
(\eps)\|T_1\|^2_{W_2^1(D)}
+c_{19}(\eps)\|S_1\|^2_{\dot W_2^1(D)}+c_{20}(\eps).$$
Now we set $X={\mathcal{H}}^s,  s\in (0,\frac14)$. In the above
discussion we have shown that the set $B$, consisting of {\em
bounded} sets, is forward invariant. Then we can apply
Theorem 4.9 above to the random dynamical system $\phi$ generated by
(\ref{rds}),  and, via the transformation (17),
get the following result to the original stochastic THC model.

\begin{theorem} \label{degree}
({\bf Finite Degrees of Freedom}) The THC model (1)-(3) has
finitely many asymptotic degrees of freedom, in the sense of having a finite
set of linearly independent continuous functionals which is
asymptotically determining in probability,  on ${\mathcal{H}}^s$ with
$0<s< 1/4$.
\end{theorem}

\section{Conclusion}

We have investigated the dynamical behavior of a random thermohaline
circulation model. We have shown that the random THC model is asymptotically described
by a random attractor (Theorem \ref{attractor}). And this system has finite degree of freedom
 in the sense of having a finite set of determining functionals in probability (Theorem \ref{degree}).


\bigskip

{\bf Acknowledgement.} We would like to thank Hongjun Gao
and Bjorn Schmalfuss for helpful discussions.


\end{document}